\documentclass[fleqn,10pt]{wlscirep}
\usepackage{subcaption,graphicx}
\title{Knot spectrum of turbulence}

\author[1]{R.~G.~Cooper}
\author[1]{M.~Mesgarnezhad}
\author[1,2]{A.~W.~Baggaley}
\author[1,2,*]{C.~F.~Barenghi}
\affil[1]{School of Mathematics, Statistics and Physics
Newcastle University, Newcastle upon Tyne NE1 7RU, UK}
\affil[2]{JQC (Joint Quantum Centre) Durham-Newcastle}
\affil[*]{carlo.barenghi@newcastle.ac.uk}

\begin{abstract}
Streamlines, vortex lines and magnetic flux tubes
in turbulent fluids and plasmas display a great amount
of coiling, twisting and linking, raising the question as to whether
their topological complexity (continually created and destroyed by
reconnections) can be quantified.  In superfluid helium,
the discrete (quantized) nature of vorticity can be exploited to
associate to each vortex loop a knot invariant called the
Alexander polynomial whose degree characterizes the topology of that vortex
loop. By numerically simulating the dynamics of a tangle of quantum 
vortex lines,
we find that this quantum turbulence always contains
vortex knots of very large degree which keep forming, vanishing
and reforming, creating a distribution of topologies which we quantify
in terms of a knot spectrum and its scaling law. 
We also find results analogous to those in the wider literature, 
demonstrating that
the knotting probability of the vortex tangle
grows with the vortex length, as for 
macromolecules, and saturates above a characteristic length, as
found for tumbled strings.
\end{abstract}

\begin{document}

\flushbottom
\maketitle

\thispagestyle{empty}

\section*{Motivation}

Tangled filamentary structures are a regular occurrence 
in physical systems ranging from turbulent fluids \cite{Ricca1996}
and magnetic fields\cite{Glatzmaier1995,Rogers2017} 
to optics\cite{Dennis2010,Irvine2008},
nematic liquid crystals\cite{Deek2013} and superfluids\cite{Barenghi2014a}.
Knowing if such structures are knotted is important;
for example, in DNA this knowledge provides valuable information on the global 
arrangement of molecules\cite{Arsuaga2005}.
In fluids, plasmas and superfluids
any increase or decrease of linking or knottiness is caused by 
vortex reconnections\cite{Kida1994,Hussain2011,Zuccher2012}, dramatic
events which are associated with viscous, resistive and acoustic 
energy losses respectively.
One would like to relate the
topology of turbulent flows to their energy, and the first step in this
direction is to precisely quantify the topology. The success
in creating a single vortex knot in controlled
laboratory conditions\cite{Kleckner2013}
has stimulated new ideas, including the use of
polynomials to analyse knotted structures. For example, it has been shown
that the decay of a single knotted vortex\cite{Kleckner2015}
can be tracked using the HOMFLYPT polynomial\cite{LiuRicca2015}. 
Unfortunately, the step from one vortex
to turbulence is still a big step.  In computational
fluid dynamics, a sample of vortex lines in a turbulent flow can be traced
and analyzed (at least in principle) to determine whether 
they are knotted or linked; however, 
moving across regions of high and low background vorticity, we may find
that, due to unavoidable numerical noise, the line does not close 
as required by the solenoidal property of vorticity.

To overcome this difficulty we note that
whereas in ordinary fluids vorticity is a continuous field of 
arbitrary shape and strength, in quantum fluids \cite{BarenghiParker2016}
such as superfluid helium and atomic Bose-Einstein condensates (BECs),
vorticity consists of individual vortex lines of fixed circulation $\kappa=h/m$
(where $h$ is Planck's constant and $m$ the mass of the relevant boson)
moving in the perfect background of inviscid potential flow.
In these quantum systems, a vortex line is an individual
topological defect (the phase of the complex wave function $\Psi$ 
vanishes on the vortex axis, hence its phase is undefined) around
which the phase changes by $2 \pi$, corresponding to the
superflow $v=\kappa/(2 \pi r)$ where $r$ is the radial distance to the
axis.  This property guarantees that (away from boundaries where 
$\vert \Psi \vert \to 0$) there is never any numerical ambiguity in 
localizing a vortex line, hence determining the linking and the knottiness
of lines.
It is therefore convenient to study the problem of the topology of
turbulence in the context of turbulent quantum fluids. This state of `quantum
turbulence' can be easily generated in the laboratory
by stirring liquid helium or atomic condensates, and takes the form of
a disordered tangle of vortex lines.

An additional motivation to consider the topology of 
quantum turbulence is that
if the superfluid is forced at length scales $D$ much larger than
the average distance $\ell$ between vortex lines 
\cite{Barenghi2014b,Laurie2012,BSB2016} so that enough k-space
is available, then,
in this wide range between $D$ and $\ell$
the quantum turbulent flow 
displays the same velocity statistics\cite{Baggaley-stats,LaMantia-stats}
and the same\cite{Salort2010} power-law distribution of kinetic energy 
(the Kolmogorov law) which is observed in ordinary (classical) turbulence. 
In this particular regime, quantum turbulence therefore
represents the `skeleton' of classical 
turbulence (other regimes of quantum turbulence exist
which do not display the Kolmogorov law and lack an energy cascade\cite{BSB2016}).

Thus motivated, our aim is to explore numerically the topology 
of quantum turbulence. 
At this early stage of investigation, the precise state of quantum turbulence 
which is examined must be chosen on the ground of its simplicity 
and  experimental availability. 

As in classical fluids, a quantum vortex
line is either a closed loops or it terminates at a boundary. To avoid
any ambiguity or arbitrariety in characterizing the linking between two
vortex lines, we restrict our
attention to turbulence in an infinite domain,
away from boundaries (assuming zero velocity at infinity), thus guaranteeing
that all vortex lines are in the shape of closed loops.  
A further simplification
is to restrict our study to statistical steady state regimes, such that
forcing and dissipation are balanced, flow properties 
fluctuate around mean values, and the memory of the
initial condition is lost. 

These considerations lead us to consider experiments in which 
quantum turbulence is    
generated at the centre of the experimental cell away from boundaries at
sufficiently high temperatures that the mutual friction between the 
vortex lines and the thermal excitations (the normal fluid)  provides
a mechanism to inject (extract) energy into (from) the vortex lines; 
this temperature regime is also the most common in experiments. 
To sustain turbulence at the centre of the sample, the liquid helium 
must be stirred by suitably positioned small oscillating 
objects \cite{VinenSkrbek2014}, such as forks \cite{Schmoranzer2016},
grids \cite{Davis2000,Bradley2011}, or wires \cite{Bradley2004},  
or by focusing 
ultrasound\cite{SchwarzSmith1981,Milliken1982}; in atomic BECs, 
the analog set-up would be to create a localized
tangle of vortex lines by stirring a large condensate with a laser spoon.

\begin{figure}[!h]
\includegraphics[angle=0,width=1.00\textwidth]{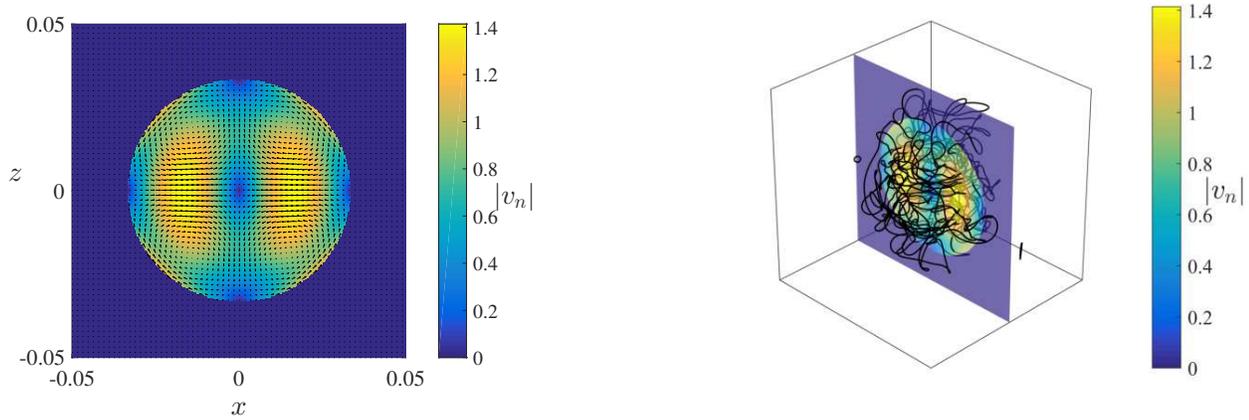}
\caption{
Left: Instantaneous Dudley-James flow: plot of $\vert \mathbf{v}_n \vert$ vs
$x,z$ at $y=0$ in the region $-0.05 < x,y <0.05~\rm cm$ with superimposed
arrowplots.
Right:
Instantaneous three-dimensional snapshot of the vortex tangle superimposed
to the magnitude $\vert \mathbf{v}_n \vert$ of the driving Dudley-James 
flow on the $x,z$ plane at $y=0$. The cube $-0.05 < x,y,z < 0.05~{\rm cm}$
around the vortex tangle is for visualization only 
(the simulation is performed in an infinite domain).
}
\label{fig1}
\end{figure}

Since we are not interested in modelling the detailed action 
of a specific oscillating object on the liquid helium but only in 
feeding/extracting energy into/from the vortex tangle in a localised 
region, we model the normal fluid's velocity field using a relatively 
simple, localised, time-dependent flow. Such flow is usually laminar
(as Reynolds numbers based on the viscosity of the normal fluid
and the velocity/size of small oscillating objects are never large).
We choose 
a Dudley-James flow \cite{DudleyJames1989} within a spherical region
of radius $D=0.03~\rm cm$ and zero outside (see Fig.~\ref{fig1}(left)
and Methods for details).
Unlike the random waves used in a preliminary investigation
\cite{Mesgarnezhad2018} which focused on the superfluid helicity, 
the Dudley-James flow is incompressible, hence more realistic.
Dudley-James flows have been used in MHD to study the kinematic dynamo 
(the growth and nonlinear saturation of magnetic field given a prescribed
velocity field),
a problem which is (at least in spirit) similar to ours (the growth and
nonlinear saturation of a tangle of quantum vortex lines given a prescribed
normal fluid velocity field). 
The initial condition of each numerical simulation consists of a few random
seeding vortex rings. The evolution of this initial condition into 
a turbulent tangle of vortex lines in a statistical steady-state is simulated
using the Vortex Filament Method (see Methods for details).

\section*{Results}

At the beginning of the time evolution,
the length of the initial vortex configuration increases
rapidly as energy is fed into the vortices by the normal fluid via
the Donnelly-Glaberson (DG) instability\cite{Tsubota2003}.
The DG instability manifests itself as growing Kelvin waves - helical displacements
of the vortex lines away from their initial position - and 
occurs if the component
of the normal fluid velocity parallel to a vortex line exceeds a critical
value.
Closely packed vortex lines interact, stretching and rotating
around each other, and reconnect when they 
collide\cite{Koplik1993,Bewley2008}.
The relaxation of each reconnection cusp releases more Kelvin waves
\cite{Fonda2014}, and a turbulent tangle of vortex lines is quickly
created. 

\begin{figure}
\centering
\includegraphics[width=0.6\linewidth]{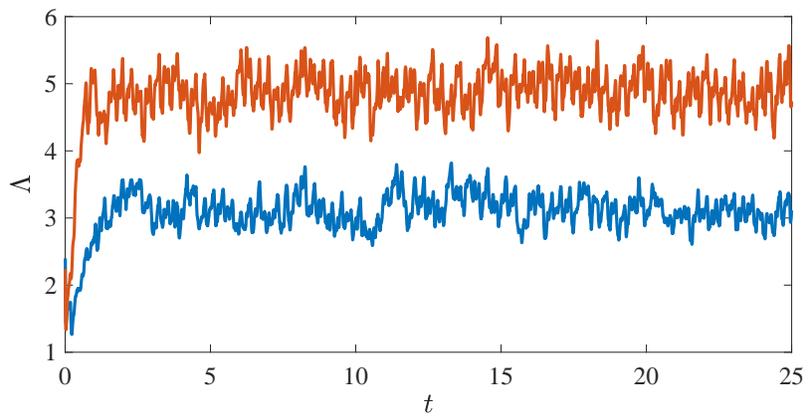}
\caption{Time evolution of the vortex line length $\Lambda$ (cm)
in the two numerical simulations. 
The lower/upper (blue/red) lines correspond to normal fluid drives 
of $v_f=4.75~\rm{cm/s}$ and $v_f=5.25~\rm{cm/s}$ respectively.
}
\label{fig2}
\end{figure}

After the initial transient, the growth of the vortex line length 
is balanced by dissipation and the vortex tangle settles to a
statistical steady-state in which the total vortex length fluctuates
around a mean value which depends on the driving normal fluid velocity,
as shown in Fig.~\ref{fig2}. This happens, firstly because
the friction with the normal fluid damps out the Kelvin waves\cite{BDV1985}; 
secondly because the vortex loops which escape the central region 
(where the normal fluid stirs the turbulence) shrink and vanish due to 
the friction with the stationary normal fluid in the outer region 
(no DG instability takes place there).
The vortex tangle, therefore, remains localised in the region of
the driving normal flow, as shown
in Fig.~\ref{fig1}(right).

\begin{figure}
\centering
\includegraphics[width=1.20\linewidth]{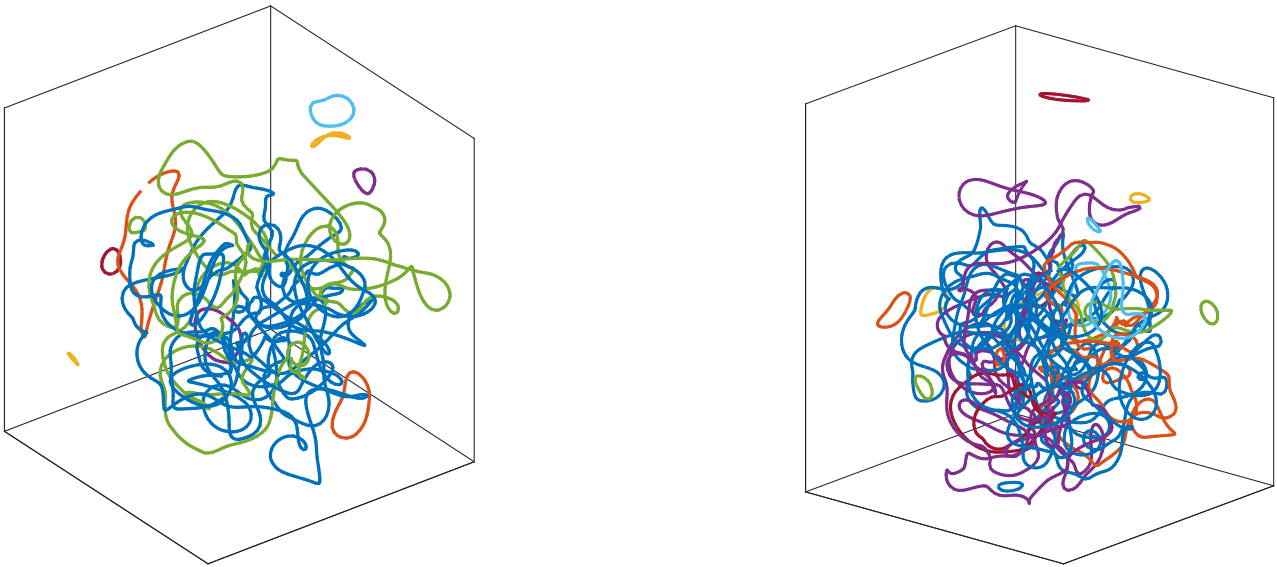}
\caption{Snapshots of the vortex configuration for the lower (left) 
and higher (right) normal fluid drives
$v_f=4.75~\rm{cm/s}$ and $v_f=5.25~\rm{cm/s}$ respectively
at $t=20$s in the saturated steady-state regime. 
Different colours are used to identify distinct vortex lines 
in the two snapshots.  The cubes around the vortex tangles are for
visualization only (the simulations are performed in an infinite domain).
}
\label{fig3}
\end{figure}

Snapshots of the vortex configurations for the two different levels of 
normal fluid drive are plotted in Fig.~\ref{fig3}: the larger the
driving normal fluid velocity, the more intense the turbulence (its
intensity is traditionally quantified in experiments in terms of the
vortex line density $L$, the length of vortex line per unit volume).
At temperatures lower than the results presented here, 
the turbulent tangles shown in Fig.~\ref{fig3} would be 
surrounded by a cloud of small vortex loops which rapidly fly 
away\cite{BarenghiSamuels2002}; at the temperatures which we consider here
these small loops are destroyed by friction. In either case,
by either shrinking and vanishing (at intermediate and high temperatures) 
or by rapidly flying away (at low temperatures)
these vortex loops represent an energy sink at short length scales.
In this way a balance is reached between drive and dissipation, resulting 
in fluctuations of the vortex length and the 
energy about mean values.

\begin{figure}
\centering
\includegraphics[width=0.60\linewidth]{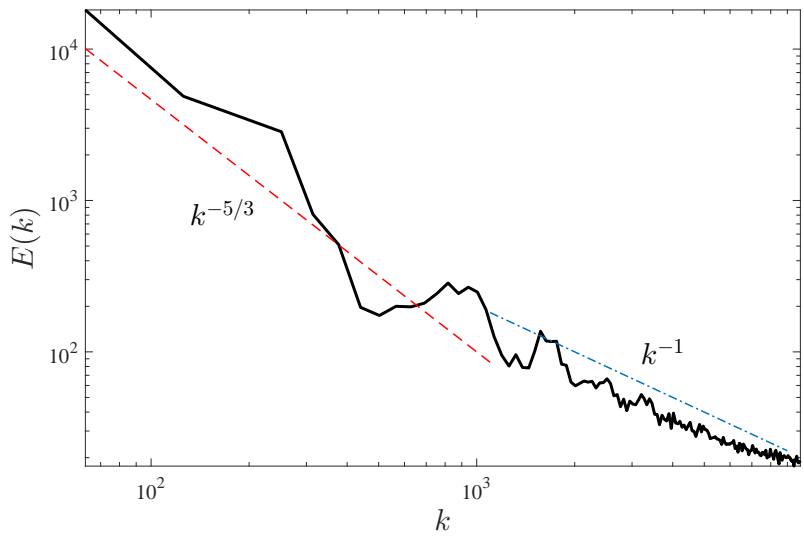}
\caption{
Instantaneous energy spectrum $E(k)$ (arbitrary units)
of the superfluid velocity at $t=24~\rm s$ (in the 
saturated regime) for $v_f=5.25~\rm cm/s$ drive
plotted vs wavenumber $k$ ($\rm cm^{-1}$).
The red and blue dashed lines are guides to the eye to
indicate the $k^{-5/3}$ and the $k^{-1}$ scaling slopes respectively.
The crossover between the two behaviours corresponds to the
average intervortex distance $\ell$.
}
\label{fig4}
\end{figure}

To demonstrate in a quantitative way that our vortex configurations
are turbulent, Fig.~\ref{fig4} shows the kinetic
energy spectrum $E(k)$ (where $k$ is the magnitude of the three-dimensional
wavenumber $\mathbf{k}$) in the saturated regime.
It is apparent that the energy injected at the large length scale $D$
cascades to smaller length scales (larger wavenumbers $k$).
As the total injected energy is balanced by friction,
most energy remains contained in the length scales larger than the
average intervortex distance $\ell$. In this region 
$k_D < k < k_{\ell}$ of k-space
(where $k_D=2 \pi/D$ and $k_{\ell}=2 \pi/ \ell$), 
we find that
the energy spectrum is consistent with the famous Kolmogorov
scaling $E(k) \sim k^{-5/3}$ of classical homogeneous isotropic
turbulence. We stress that we should not expect a precise
$k^{-5/3}$ scaling in our simulations: our turbulent flow is not
homogeneous and only one decade of k-space is available. The crossover
from the Kolmogorov $k^{-5/3}$ scaling to the $E(k) \sim k^{-1}$ 
scaling  which is typical of individual vortex lines occurs at
approximately $k \approx 10^3~\rm cm^{-1}$, which is consistent with 
$k_{\ell} \approx 1300~\rm cm^{-1}$ estimated from the vortex 
line density $L=\Lambda/V$ where $V \approx 4 \pi D^3/3$ 
is the volume of the turbulent region and 
$\ell \approx L^{-1/2}\approx 4.8 \times 10^{-3}~\rm cm$ is the
average intervortex spacing corresponding to the numerical
simulation at large drive ($\Lambda \approx 5~\rm cm$).

Our technique to quantify the topology of the turbulence
is based on a knot invariant: the Alexander polynomial\cite{Alexander1928}.
The polynomial  
$\Delta(\tau)=a_0 + a_1 \tau + \cdots + a_{\nu} \tau^{\nu}$
of degree $\nu$ with integer coefficients $a_0, \cdots , a_{\nu}$ is assigned 
to a knot type.  For example, a vortex ring has Alexander
polynomial $\Delta(\tau)=1$ which identifies the {\it unknot}; therefore
any closed vortex loop which can be continuously (i.e. without reconnections)
deformed into a ring corresponds to $\Delta(\tau)=1$. The simplest
non-trivial  knot is the {\it trefoil knot} which has Alexander polynomial
$\Delta(\tau)=1 -\tau + \tau^2$ of degree $\nu=2$. 
In general, the higher the degree $\nu$ of
the Alexander polynomial, the more complex the knot type of the vortex loop. 

The instantaneous vortex configuration consists of $N$ vortex loops
${\cal L}_i$ ($i=1, \cdots N$), where $N$ changes with time as 
vortex reconnections continually merge and split vortex loops. After the initial transient,
$N$ fluctuates about a mean value. In this statistical steady state regime,
at any given time $t$, we numerically
determine the Alexander polynomial 
of each loop ${\cal L}_i$ ($i=1, \cdots N$) and call $\nu_i$ its degree.
We stress that our simulations are not stochastic:
the only source of randomness in the tangle's topology arises 
from the turbulent fluctuations.
Data are collected over time (in the steady-state regime) and cumulated 
to generate statistics for the analysis.

Our initial focus is what determines the probability that a given loop is knotted.
Perhaps unsurprisingly, this fundamental question has been addressed in studies of
other physical systems that contain knots.
For example, Arsuaga {\it et al.} \cite{Arsuaga2002} performed Monte Carlo simulations 
of random knotting in confined volumes, which were directly compared to the visualisation of 
DNA molecules. Their simulations suggested a linear relationship between the length of 
a loop and its probability of being knotted, which we shall denote $P_k$. 
A more recent study by Raymer \& Smith \cite{Raymer2007} performed experiments where strings 
of different lengths were tumbled inside a box. They found that $P_k$ grows rapidly with
the length of the string, and saturates to some limiting value above 
a characteristic string length. Interestingly, the limiting value which they found was 
substantially less than 1, which they attributed to the finite agitation time and the
stiffness of the strings.  

\begin{figure}
\centering
\includegraphics[width=0.45\linewidth]{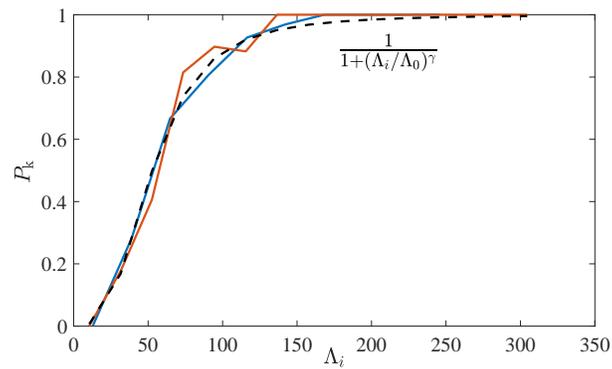} 
\caption{Probability $P_k$ that a vortex loop is knotted plotted
vs the loop's length $\Lambda$. We have fitted data from both simulations 
($v_f=4.75~\rm cm/s$ and $5.25~\rm cm/s$, red and blue curves respectively)
with the same sigmoidal curve (black dashed curve)
with fitting parameters $\Lambda_0=53~\rm cm$, $\gamma=-3.1$. 
Bin widths are taken to be approximately $20$cm.}
\label{fig5}
\end{figure}

\begin{figure}
\centering
\includegraphics[width=0.45\linewidth]{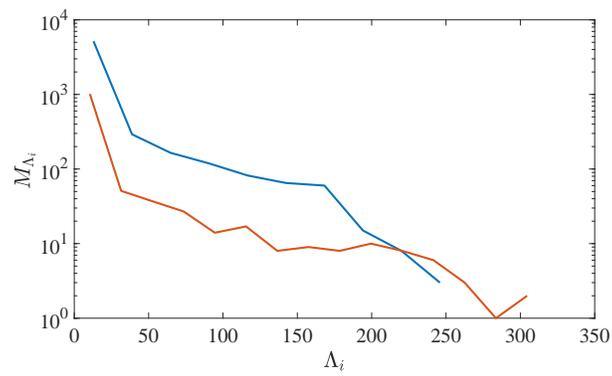}
\caption{
The number of knots, $M_{\Lambda_i}$ within each of the bin widths,
$\Lambda_i$, used in Fig.~\ref{fig5}.}
\label{fig6}
\end{figure}

We now proceed with the analysis of our results.
Fig.~\ref{fig5} displays $P_k(\Lambda)$, the probability that a
loop of length $\Lambda$ is knotted, calculated in the saturated
steady-state regimes for both small and large
normal fluid drives considered 
in this study. We find that $P_k(\Lambda)$ is independent of the
normal fluid stirring (at least for the values considered here) 
and depends only on the loop length. Following Raymer \& Smith
\cite{Raymer2007}, we fit a sigmoidal curve to the data of the form 
$P_k(\Lambda)=(1+(\Lambda/\Lambda_0)^{\gamma})$. 
Fitting our data using
$v_f=4.75 {\rm cm/s}$, we find $\Lambda_0=53~\rm cm$ and $\gamma=-3.1$,
in fair agreement with $\gamma=-2.9$ reported by
Raymer \& Smith \cite{Raymer2007}, suggesting that the
knotting probabilities of quantised vortices moving under reconnecting
Biot-Savart dynamics and tumbled strings are not unrelated. 
To show that our results are statistically robust in the limit of 
large $\Lambda_i$,   
Fig.~\ref{fig6} displays the number of knots, $M_{\Lambda_i}$, 
within each of the bins of the histogram.
Notice that in general there are at least 10 vortex loops
within a bin, until $\Lambda_i>250$ for the higher drive simulation 
which is well into the asymptotic regime of $P_k=1$.

\begin{figure}
\centering
\includegraphics[width=0.6\linewidth]{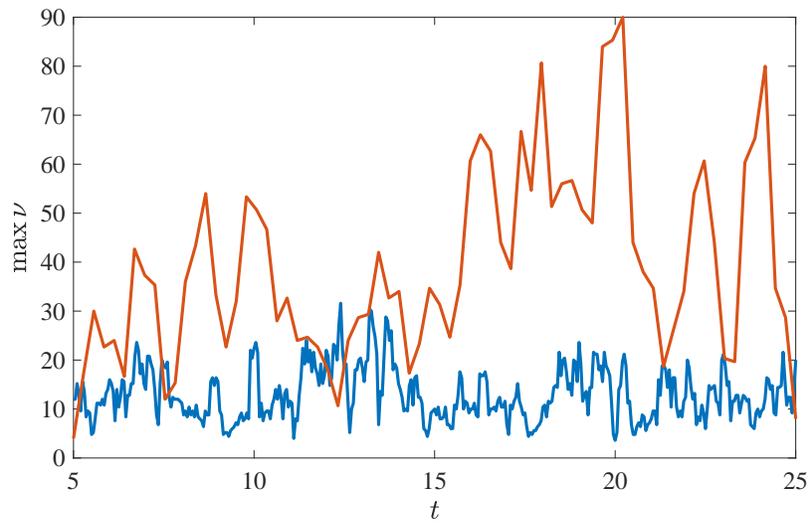}
\caption{
Time series of the largest degree of Alexander polynomial in the vortex configuration. 
The upper line is for $v_f=4.75~\rm cm/s$, and the lower line for $v_f=5.25~\rm cm/s$. }
\label{fig7}
\end{figure}

\begin{figure}
\centering
\includegraphics[width=0.6\linewidth]{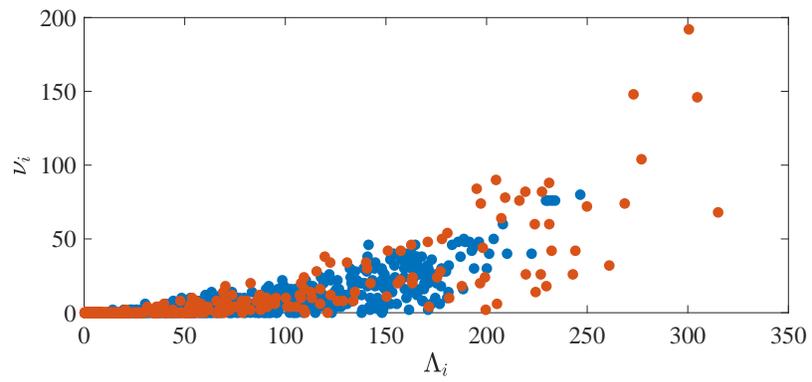}
\caption{
The degree $\nu_i$ of the Alexander polynomial of each vortex loop 
plotted against its length $\Lambda_i$ (in cm) for the lower normal fluid 
drive ($v_f=4.75~\rm cm/s$, blue symbols) and the higher drive 
($v_f=5.15~\rm cm/s$, red symbols). }
\label{fig8}
\end{figure}

We now move to consider the topological
complexity of individual vortex loops
and how it depends on the normal fluid drive and the loop's 
length. As we have discussed, at any 
instant of time, the vortex tangle contains knots, and 
the most complex knot has the largest value of $\nu$. 
Figure \ref{fig7} shows the time series of the maximal 
degree of Alexander polynomial present in the vortex tangle, revealing
that at all times 
the vortex tangle contains a significant topological complexity 
 - we never observe that the tangle is simply
a collection of unknots. It must be stressed that the very knotted 
vortex structures which
we detect are unstable: they are continually broken down by vortex 
reconnections, but they keep reforming by further reconnections.
The natural question is whether the complexity depends on the driving
normal fluid velocity. To answer the question we show a scatter-plot 
of $\Lambda_i$ vs $\nu_i$, see Fig.~\ref{fig8}. It is apparent that
complexity is strongly related to the loop's length, 
which is perhaps unsurprising. However, as for $P_k(\Lambda)$, 
the underlying functional relationship 
between $\nu_i$ and $\Lambda_i$ appears independent of $v_f$, 
so the additional complexity which is evident in Fig.~\ref{fig8} 
is simply due to a stronger drive's propensity to 
generate longer loops (larger values of $\Lambda_i$).

\begin{figure}
\centering
\includegraphics[width=0.6\linewidth]{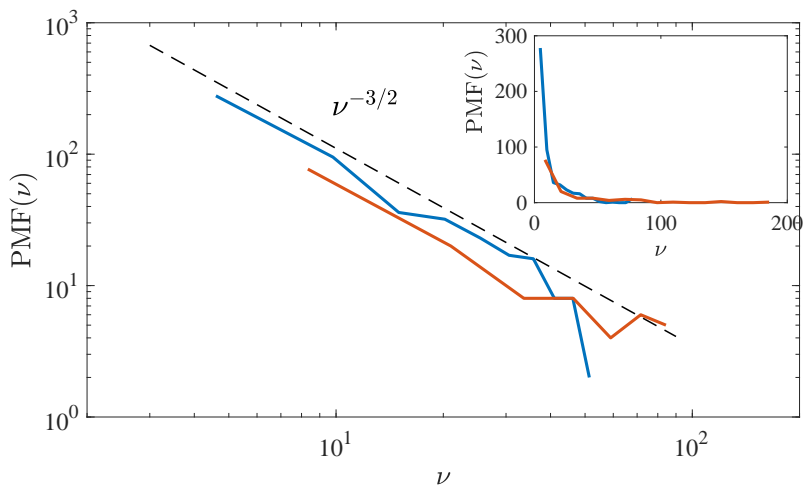}
\caption{ Probability mass functions (PMFs) of the degree 
of the Alexander polynomial $\nu$ plotted on a log-log scale (main
graph) and linear-linear scale (inset). The logarithmic scale 
suggests $PMF(\nu) \sim \nu^{-3/2}$, as shown by the black dashed line.}
\label{fig9}
\end{figure}

Finally, attempting to gain a probabilistic understanding of the 
topological complexity of quantum turbulence which in the future 
could be compared to the distribution of knots in DNA, polymers, etc,
we consider the distribution of $\nu_i>0$. Fig.~\ref{fig9} shows the
Probability Mass Function (PMF) (normalized histogram of discrete
variable) of the degrees of Alexander polynomials:
${\rm PMF}(\nu_i)$ vs $\nu_i$.
We find two striking results: firstly that the PMFs are again 
independent of the normal fluid drive, and secondly that there appears 
a distinct scaling 
${\rm PMF}(\nu) \sim \nu^{-3/2}$ which we can call the `knot spectrum'
of turbulence. 

\section*{Discussion}

We have exploited the key property of quantum fluids - the discrete
nature of vorticity - to quantify the topology of a small region
of quantum turbulence (a vortex tangle) away from boundaries in a statistical 
steady-state regime. 
Such inhomogeneous vortex configuration can be realized in experiments in which
turbulence is driven by small oscillating 
objects\cite{VinenSkrbek2014,Schmoranzer2016,Davis2000,Bradley2011,Bradley2004}
or by focussing ultrasound\cite{SchwarzSmith1981, Milliken1982}. 
The temperature regime which we have chosen is typical of many $^4$He
experiments.

We have found that the probability that a vortex loop is knotted
increases with the loop's length as for random knots studied in the
context of DNA and macromolecules\cite{Arsuaga2005}, 
and saturates above a characteristic length
as for tumbled strings \cite{Raymer2007}, despite the
very different physical
mechanisms of agitation of these systems
(respectively Brownian motion, mechanical chaos, 
the Biot-Savart law of Eulerian fluid dynamics).
We have also found that, at any instant of time, the vortex tangle contains
a distribution of vortex knots. We have quantified the
topological complexity of these knots in terms of the degree of
the Alexander polynomial which we numerically associate to each vortex loop.
Surprisingly, the turbulence always contains some very long vortex
loops of great topological complexity, and the distribution of this
complexity (measured by the degree of the Alexander polynomials) displays
a scaling law, or knot spectrum.

With more computing power available, future work will consider larger, denser
vortex tangles at different temperatures and drives, including the
zero temperature limit, with the aim of determining
the dynamical origin of the characteristic length
for the knotting probability and of verifying if the knot spectrum scaling is
universal or not.  
One should also extend this initial study
to flows in larger periodic domains (where the turbulence can be driven
to a more accurate classical Kolmogorov regime), and to flows within 
channel boundaries, where tools from braid theory may be used.  
Finally, it would be interesting to 
perform a similar analysis using HOMFLYPT polynomials.

\newpage

\section*{Methods}

\subsection*{Superfluid:}
We use the Vortex Filament Model (VFM) of Schwarz\cite{Schwarz1988}.
The method is based on the observation that the
average separation between vortex lines
in typical superfluid helium ($^4$He) experiments is 
$\ell \approx 10^{-5}$ or $10^{-6}~\rm m$, which is many
orders of magnitude bigger than the radius of the vortex core,
$a_0 \approx 10^{-10}\rm m$.
We can therefore model superfluid vortex lines as closed space-curves 
${\bf s}(\xi,t)$ of thickness which is infinitesimal to any other length
scale of the flow, where $t$ is time and
$\xi$ is arc-length. The velocity of a vortex line at
the point ${\bf s}$ is given by Schwarz's equation\cite{Schwarz1988}

\begin{equation}
\frac{d{\bf s}}{dt}={\bf v}_{self} 
+ \alpha {\bf s'} \times ({\bf v}_n - {\bf v}_{self}) 
- \alpha'{\bf s'} \times [ {\bf s'} \times ({\bf v}_n - {\bf v}_{self}) ],
\label{eq:Schwarz}
\end{equation}

\noindent
where ${\bf s}'=d{\bf s}d\xi$ is the unit tangent vector to the curve
at the point ${\bf s}$,  ${\bf v}_n$ is the velocity of the normal 
fluid at $\bf s$, 
and $\alpha$ and $\alpha'$
are small dimensionless temperature-dependent
friction coefficients\cite{BDV1983,DB1998}.
The self-induced velocity of the vortex line
at the point $\bf s$ is given by the 
Biot-Savart law\cite{Saffman1992}

\begin{equation}
{\bf v}_{self}({\bf s})=
-\frac{\kappa}{4\pi}\oint_{\mathcal{L}} \frac{({\bf s-r}) \times {\bf dr}}{\left|{\bf s-r}\right|^3},
\label{eq:BS}
\end{equation}

\noindent
where the line integral extends over all vortex lines: 
${\cal L}=\cup_{i=1}^{N} {\cal L}_i$.
In the low-temperature limit $T \to 0$, the friction
coefficients vanish,  vortex lines are simply advected by the flow
which they generate, and Schwarz's equation reduces to
$d{\bf s}/dt = {\bf v}_{self}(\bf s)$ in agreement with the
classical Helmholtz's Theorem. 
In practice, the zero-temperature limit is a good approximation
for for $T < 1~\rm K$; 
At higher temperatures,
the normal fluid fraction is not negligible and the friction terms
must be included in Schwarz's equation.
In our numerical simulations
we choose the temperature $T=1.9~\rm K$ which is typical of many
experiments and corresponds to $\alpha=0.206$
and $\alpha'=0.00834$; at this temperature, the superfluid and normal fluid
fractions are respectively $\rho_s/\rho=0.58$ and $\rho_n/\rho=0.42$, where
$\rho_s$ is the superfluid density, $\rho_n$ the normal fluid density
and $\rho=\rho_s+\rho_n$ the total density of liquid helium.

Our numerical model uses a variable Lagrangian 
discretization along the vortex lines\cite{Schwarz1988} in which
the density of discretization points depends on the local 
curvature.  The Biot-Savart integral in equation~(\ref{eq:BS}) 
is de-singularised 
in a standard way\cite{Schwarz1988} based on the vortex core cutoff $a_0$,
and the procedure for vortex reconnections is implemented algorithmically
\cite{Schwarz1988,Baggaley2012}.
The initial condition consists of
40 randomly oriented loops of radius varying according to a normal 
distribution and with an average number of 200 discretization points on
each; the initial rings are located at the centre of the infinite 
computational domain.
The vortex  configuration remains localised in a finite volume 
due to our choice of normal fluid (see below) and temperature, 
it is important to emphasise that the superfluid velocity 
is computed in unbounded space, i.e. without the presence of
external boundary conditions.
The total length of the vortex configuration is 
\begin{equation}
\Lambda=\oint_{\cal L} d\xi,
\end{equation}
and similar integrals over distinct loops (${\cal L}_i$) determine 
the length of the $i^{\rm th}$ loop, $\Lambda_i$. Finally, the superfluid energy
spectrum $E(k)$ is defined by
\begin{equation}
E=\frac{1}{V}\int_V \frac{\rho_s}{2} \vert \mathbf{v}_s \vert^2 dV=
\int_0^{\infty} E(k) dk.
\end{equation}

\subsection*{Normal fluid}
Following the classical turbulence literature,
quantum turbulence is usually studied
in statistically stationary, homogeneous regimes in
three-dimensional periodic domains\cite{Baggaley2011fluct}. 
However, periodic boundary conditions complicate the definition of the
topological properties of the vortex tangle.
Fortunately there exist experimental techniques to generate
turbulence away from boundaries in $^4$He or $^3$He-B 
by oscillating objects \cite{VinenSkrbek2014} such as
forks \cite{Schmoranzer2016}, grids \cite{Davis2000,Bradley2004},
wires \cite{Bradley2004}, or
by focussing ultrasound \cite{SchwarzSmith1981,Milliken1982}.
The microscopic details of the vortex nucleation (e.g. the
implosion of cavitating bubbles which create vortex rings 
\cite{Berloff2004} or the interaction with the unavoidable
microscopic irregularities of a hard boundary\cite{Stagg2017})
are not relevant here - we are only concerned on what happens to
the vortex lines once they are injected into the fluid.

At finite temperatures, a steady state of quantum turbulence can be maintained by the injection of energy from the normal fluid,
whose velocity field we denote ${\bf v}_n$. Hence, to study the topology of a vortex tangle in a statistically steady state of turbulence, but confined to a finite volume of space, we can make use of a confined normal fluid flow. 
In a previous study \cite{Mesgarnezhad2018}, we made use of random waves
modulated by an exponential window so that the velocity field decayed rapidly like $e^{-|\mathbf{x}|^2}$. Whilst this is mathematically convenient, 
physically it is unrealistic as the normal fluid's velocity field 
is no longer solenoidal.
Here we make use of a Dudley-James flow \cite{DudleyJames1989}, which is 
well known in the context of dynamo theory and provides a convenient analytic 
velocity field which is both confined and solenoidal. Importantly,
the flow is sufficiently complex to support a magnetic dynamo, 
and here we shall show it can act effectively as a quantised vorticity dynamo.
Using radial coordinates $(r,\theta,\phi)$, we assume the following form 
for the velocity field:
\begin{equation}
\mathbf{v}_n(r,\theta,\phi)=\sum_{l,m} \mathbf{t}_l^m + \mathbf{s}_l^m,
\end{equation}
where,
\begin{equation}
 \mathbf{t}_l^m=\nabla \times \hat{\mathbf{r}} t_l^m Y_l^m(\theta,\phi), \qquad  \mathbf{s}_l^m=\nabla \times \nabla \times \hat{\mathbf{r}} s_l^m Y_l^m(\theta,\phi), \quad -l \le m \le l.
\end{equation}
Here we consider the following Dudley \& James flow:
\begin{equation}
\mathbf{v}_n=\mathbf{t}_2^0+\epsilon \mathbf{s}_2^0,
\end{equation}
with
\begin{equation}
t_2^0=s_2^0=r^2 \sin(\pi r/D).
\end{equation}
To introduce time dependence, we take for the normal fluid velocity the
form
\begin{equation}
\mathbf{v}_n=
\begin{cases}
A(t) \mathbf{t}_2^0+B(t) \epsilon \mathbf{s}_2^0, \quad r \le D\\
0,\quad r >D,
\end{cases}
\end{equation}
with $A(t)=v_f \sin \omega t$ and $B(t)=v_f \cos \omega t$; 
in all simulations presented we take $\epsilon=0.2$, 
and $\omega=20 \rm{s}^{-1}$. We consider two levels of `drive': 
$v_f=4.75 \rm{cm/s}$ and $v_f=5.25 \rm{cm/s}$; the value of $v_f$
sets the amount of energy injected from the normal fluid into the
superfluid vortex lines, hence the  vortex line length in the steady-state 
regime after the initial transient.

\subsection{Alexander polynomial}

In order to quantify the topological complexity of the vortex loop
${\cal L}_j$ we determine its Alexander polynomial

\begin{equation}
\Delta_j(\tau)=a_0+a_1\tau+...+a_{\nu_j}\tau^{\nu_j}.
\end{equation}

\noindent
The degree $\nu_j$ of the polynomial quantifies
 the complexity of the
loop. 
For example, the Alexander polynomial of the trivial {\it unknot}
is $\bigtriangleup (\tau)=1 $ hence its degree is $\nu=0$. 
The simplest nontrivial knot is the {\it trefoil} ($3_1$) knot, 
which has Alexander polynomial
$\bigtriangleup (\tau)=1-\tau+\tau^2$, hence its degree is $\nu=2$.
Any vortex loop which has an Alexander polynomial of degree $\nu > 0$ 
is knotted 
(however the converse is not necessarily true: a long-standing 
problem of knot theory is the lack of a unique method of distinguishing 
knots from each other; in particular, the Alexander polynomial is not 
unique to a particular knot type. For example there exist knots which 
have the same Alexander polynomial as 
the {\it unknot}~\cite{DesCloizeaux1979}, so the fact that a vortex 
loop has an Alexander polynomial of degree $\nu=0$ does not necessarily 
imply that it is an {\it unknot}.)

\begin{figure}
\centering
\includegraphics[angle=0,width=0.80\textwidth]{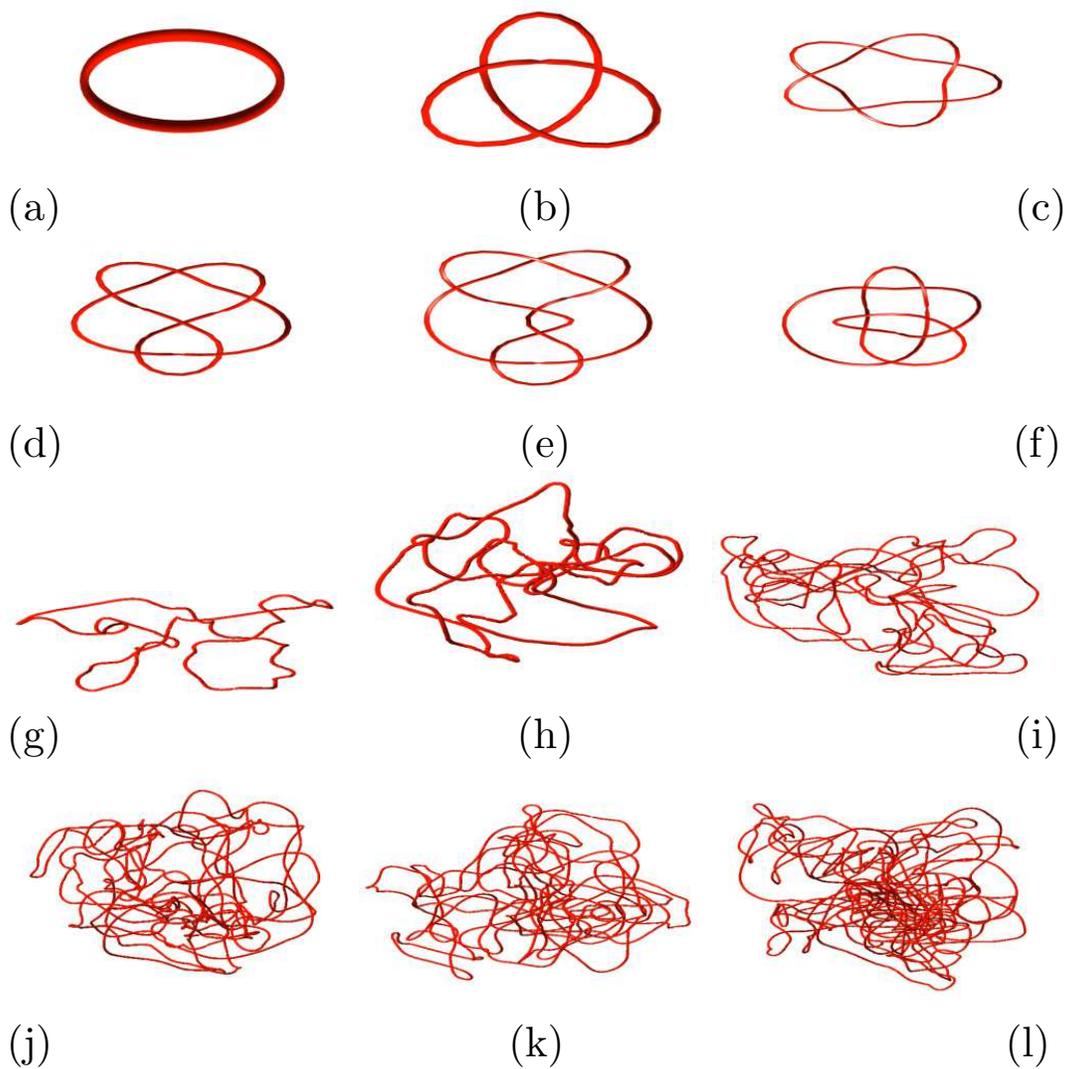}
\caption{Examples of knots and the degrees of their Alexander polynomials. 
First (top) row, from left to right: $\nu=0$, $\nu=2$ and $\nu=4$;
Second row, from left to right: $\nu=4$, $\nu=4$ and $\nu=4$;
Third row, from left to right: $\nu=0$, $\nu=8$ and $\nu=46$;
Fourth (bottom) row, from left to right: $\nu=82$, $\nu=108$ and $\nu=232$.
}
\label{fig10}
\end{figure}

Fig.~\ref{fig10} shows a selection of standard  
and numerically simulated knots with the degree $\nu$ of
their respective Alexander polynomials.
The six knots on the first two rows are standard knots; 
the {\it unknot}, the $3_1$ knot (also known as the trefoil knot), 
the $5_1$ knot (Solomon's Seal),
the, $6_2$ knot, the $7_5$ knot and the $8_{21}$ knot. 
It is apparent that the degree $\nu$ increases with the
knots' complexity. The first (left) knot of the third row,
despite its complex appearance, 
has Alexander polynomial with 
degree $\nu=0$, and indeed, by untwisting it in visible locations,
it can be easily manipulated
into an unknot. The remaining knots are obtained from numerical 
simulations of vortex lines.

Following Livingstone\cite{Livingstone1993}, 
we compute the Alexander polynomial of a vortex 
loop by labelling segments of a loop between
under-crossings when projected into a plane, followed by assigning 
coefficients to the relevant entries of
a matrix for each segment, and then finding the determinant of the matrix
with any single row and column removed \cite{Livingstone1993}.
The numerical algorithm,  described in R.G. Cooper's MMath thesis
(https://www.jqc.org.uk/publications/theses/),  was tested against
all the knots of the Rolfsen knot table 
(http:$//$katlas.org/wiki/The\_Rolfsen\_Knot\_Table). 
To test knots with very
large number of crossings we applied rotations to numerically simulated 
knots: a rotation changes the number of crossings onto a 2D plane, 
hence the matrix to determine the Alexander polynomial. 
A third test consisted in
numerically deforming part of a loop adding `false' crossings (which
could be untwisted easily if one had the knot in one's hands): these `false'
crossings change the size of the matrix used to compute the Alexander
polynomial.


\subsection*{Data Avaliability}

The datasets generated and analyzed during the current study 
are available from Newcastle University's repository.


\section*{Acknowledgements}

C.F.B. acknowledges the support of EPSRC grant number RES/0581/7397.

\section*{Author contributions statement}

C.F.B. conceived this numerical experiment; 
A.W.B. wrote and tested the vortex code; R.G.C. wrote and tested the 
Alexander polynomial code; R.G.C. and M.M. performed the numerical simulations;
R.G.C. analyzed the results; C.F.B., A.W.B. and R.G.C. wrote the manuscript.


\section*{Competing financial interests}

The authors have no competing financial or non-financial interests.


\clearpage

\clearpage

\clearpage

\clearpage

\clearpage

\clearpage

\clearpage

\clearpage

\clearpage

\clearpage

\end{document}